\documentclass[12pt]{article}
\input psfig.tex
\begin{document}

\def\ov{\overline}
\def\ra{\rightarrow}
\def\klc{\kappa_{L}^{CC}}
\def\krc{\kappa_{R}^{CC}}
\def\kln{\kappa_{L}^{NC}}
\def\krn{\kappa_{R}^{NC}}
\def\zbb{{\mbox {\,$Z$-$b$-${b}$}\,}}
\def\ttz{{\mbox {\,$t$-${t}$-$Z$}\,}}
\def\ttzz{{\mbox {\,$t$-${t}$-$Z$-$Z$}\,}}
\def\ttww{{\mbox {\,$t$-${t}$-$W$-$W$}\,}}
\def\wwtt{{\mbox {\,$W^+W^-t{\bar t} \;$}\,}}
\def\tbw{{\mbox {\,$t$-${b}$-$W$}\,}}
\def\bea{\begin{eqnarray}}
\def\ena{\end{eqnarray}}
\def\beq{\begin{equation}}
\def\enq{\end{equation}}

\setcounter{footnote}{1}
\renewcommand{\thefootnote}{\fnsymbol{footnote}}

%=========================Title Page=========================
\begin{titlepage}

{\small
\noindent
{March 1999} \hfill {MSUHEP-90315}}

\vspace{2.0cm}

\centerline{\Large\bf Analysis of $tbW$ and $ttZ$ couplings}
\centerline{\Large\bf from CLEO and LEP/SLC data}

\vspace{1.2cm}
\baselineskip=17pt
\centerline{\normalsize  
F. Larios$^a$, M. A. P\'erez$^b$  and C.-P. Yuan$^c$}
 
\vspace{1.0cm}
\centerline{\normalsize\it
$^a$Departmento de F\'{\i}sica Aplicada, CINVESTAV-M\'erida}
\centerline{\normalsize\it
A.P. 73,  97310 M\'erida, Yucat\'an,  M\'exico}

\vspace{0.5cm}
\centerline{\normalsize\it
$^b$Departmento de F\'{\i}sica, CINVESTAV}
\centerline{\normalsize\it
Apdo. Postal 14-740, 07000 M\'exico}

\vspace{0.5cm}
\centerline{\normalsize\it
$^c$Department of Physics and Astronomy, Michigan State University}
\centerline{\normalsize\it
East Lansing, Michigan 48824 , USA}

\vspace{0.4cm}
\raggedbottom
\setcounter{page}{1}
\relax

\begin{abstract}
\noindent
We update the constraints on anomalous dimension four $\tbw$
and $\ttz$ couplings by using CLEO $b \ra s \gamma$ and 
LEP/SLC precision $Z$-pole data.
It is found that the data imposes very stringent bounds on
them.  Moreover, the $2 \sigma$ pull from SM predictions
of $A_{LR}$(hadrons), $A_b$ and $A_{FB}$(b) have little chance
of being explained by the strongly constrained anomalous couplings.
\end{abstract}

\vspace*{3.4cm}
\end{titlepage}
%===============end of Title Page=========================

\normalsize\baselineskip=15pt
\setcounter{footnote}{0}
\renewcommand{\thefootnote}{\arabic{footnote}}

\section{Introduction}
\label{intro}
\indent 

The mechanism of electro-weak symmetry breaking (EWSB)
is still not known and until there is
experimental observation of the scalar Higgs boson,
the generation of masses for the
$W$ and $Z$ bosons, and the fermions,
will remain a mystery.  If the mechanism that generates
fermion masses is to be related to the EWSB,
the interaction of the top quark, 
with a mass of $\sim 174$\,GeV \cite{cdfd0}
(the same order as the EWSB scale 
$v={(\sqrt{2}G_F)}^{-1/2}=246$\,GeV), may reveal 
information on the EWSB sector.

In this work we update the constraints on dimension four
anomalous couplings of the top quark with the gauge bosons
by comparing the recent NLO calculation of the $B\to X_s\gamma$ 
decay rate with the most recent CLEO and LEP/SLC data.
Furthermore, 
given that the forward-backward asymmetries of $Z\to b\bar b$
measured at LEP and SLC show a $1.8 \sigma$ deviation from the
SM prediction, we study the possibility of these anomalous
couplings to explain such deviations.

\section{Dimension Four Anomalous Couplings}
\label{couplings}
\indent

The deviations from the $\tbw$ and $\ttz$ couplings are considered
in the context of the non-linear electroweak chiral Lagrangian,
which is the most general effective Lagrangian that can describe
decoupled or non-decoupled new physics effects \cite{malkawi}.  
Assuming no new physics effects in the neutral current bottom
quark couplings, there are four coefficients that measure the
deviation from the SM third family quark (top $t$ and bottom $b$) 
and gauge boson ($W^\pm$ and $Z$)
couplings, they are defined as follows \cite{malkawi}:
\begin{eqnarray}
{\cal L}&=&
\frac{g}{2c_w}\left (1-\frac{4s_w^2}{3}+
\kappa_{L}^{\rm {NC}}\right)
\overline{t_{L}}\gamma^{\mu} t_{L}Z_{\mu}
+ \frac{g}{2c_w}\left ( \frac{-4s_w^2}{3}+
\kappa_{R}^{\rm {NC}}\right ) \ov {{t}_{R}}
\gamma^{\mu} t_{R}Z_{\mu} \nonumber \\
&&+\frac{g}{\sqrt{2}}
\left (1+\kappa_{L}^{\rm {CC}}\right ) \ov {{t}_{L}}
\gamma^{\mu} b_{L}
{W_{\mu}^+}+\frac{g}{\sqrt{2}}
\left( 1+{\kappa_{L}^{\rm {CC}}}^{\dagger}\right)
{{b}_{L}}\gamma^{\mu}t_{L}{W_{\mu}^-} \nonumber \\
&&+\frac{g}{\sqrt{2}}\kappa_{R}^{\rm {CC}}
\overline{{t}_{R}}\gamma^{\mu} b_{R}
{W_{\mu}^+}+\frac{g}{\sqrt{2}}{\kappa_{R}^{\rm {CC}}}^{\dagger}
 \overline{{b}_{R}}\gamma^{\mu} t_{R}{W_{\mu}^-} 
 \label{lagrangian} \,\, .
\end{eqnarray}
In the above equation $\kappa_{L}^{\rm {NC}}$, $\kappa_{R}^{\rm {NC}}$,
$\kappa_{L}^{\rm {CC}}$, and $\kappa_{R}^{\rm {CC}}$
parameterize possible deviations from the SM predictions~\cite{malkawi}.
($t_L$ denotes a top quark with left-handed chirality, etc.) 
In general, the charged current coefficients can be complex with the
imaginary part introducing a CP odd interaction.
The decay process $B\to X_s\gamma$ depends on the real and imaginary
parts of $\klc$ and $\krc$, although the contribution from $\klc$ is
suppressed by $m_b$.   Previous analysis of their allowed values
have shown that these couplings could be large, even of order
$1$, but in a correlated manner \cite{malkawi}.  
A similar conclusion can be drawn from the  
partial wave unitarity bounds \cite{iowa}.
In this update we show that the correlation has become so
tight that even a deviation of the SM $\tbw$ coupling of order
$5 \%$ would require a similar deviation for the $\ttz$
couplings in order to be consistent with the LEP/SLC data.

\section{The right handed $\tbw$ coupling and $b\to s \gamma$}
\indent

The latest measurement of the $B\to X_s\gamma$ branching ratio 
(Br) by CLEO collaboration \cite{cleo98} gives 
\bea
\mbox{Br}^{exp} (B\to X_s\gamma) &=& 3.15 \pm
0.35^{stat} \pm 0.41^{sys} \, , \label{bsgaexp} \\
\mbox{with} \;\; && 2.1 \leq \mbox{E}_\gamma
\leq 2.7 \nonumber  \, \mbox{GeV} \, ,
\ena
where $\mbox{E}_\gamma$ is the energy of the decay photon.
It is roughly a $20 \%$ reduction in the error and 
a $40 \%$ shifted mean value, which is closer to the SM prediction, 
as compared with the 1995 result \cite{cleo95}.
There has also been an improvement in the SM prediction,
in which the next-to-leading order (NLO) QCD corrections
have been calculated to reduce the renormalization scale
dependence \cite{chetyrkin}.

Using the recent NLO calculation we can write the
branching ratio in terms of the $C_7$ and $C_8$
coefficients at the scale of $W$ boson mass $M_W$\cite{neubert1}:
\bea
\mbox{Br}( B\to X_s\gamma ) \times 10^{4}
&&= 1.355 - 6.67 \mbox{Re}( \; C_7(M_W))-
1.22 \mbox{Re}(C_8(M_W) \; ) \nonumber \\
&&+ 5.79 |C_7(M_W)|^2 +
0.3|C_8(M_W)|^2 \nonumber \\
&&+ 2.75 \mbox{Re}( \; C_7(M_W) C^*_8(M_W) \;) \; ,
\label{bsgral}
\ena
where the numerical factors were obtained
using the pole masses of top and bottom quarks as
$m_t=174$\,GeV, $m_b=4.8$\,GeV, and the strong coupling at 
the $Z$-mass scale to be $\alpha_s=0.118$.
Furthermore, the energy of the decay photon is required to be larger
than $(1-\delta) \mbox{E}^{max}_\gamma$ with $\delta=0.125$, which 
corresponds to the experimental cut of the photon energy range.
In Eq.~(\ref{bsgral}), the magnetic and chromomagnetic dipole 
coefficients $C_7$ and $C_8$ are sensitive to the $\tbw$ coupling.
At one loop level, they receive contributions from the type of 
new physics listed in Eq.~(\ref{lagrangian}) as \cite{yama}:
\bea
C_7 (m_W) &=& 
- (1+\klc)\, \frac{1}{2 (x-1)^{4}} \left[
\frac{2x^2-3x^3}{2} \ln(x) + \frac{x-1}{12}
(8x^3+5x^2-7x ) \right] \nonumber \\
&+&  \frac{m_t}{m_b} \krc \frac{1}{2 (x-1)^{3}} \left[
\frac{2}{3} (\, 2+3x\ln(x)-\frac{x^3}{2}-\frac{3x}{2} \,)
\right. \nonumber \\
&+& \left. 
( -\frac{x^3}{2}+6x^2-\frac{15}{2}x+2-3x^2 \ln(x) \,)
\right] \, ,
\nonumber \\
C_8 (m_W) &=&
- (1+\klc) \,\frac{1}{2 (x-1)^{4}} \left[
\frac{3}{2} x^2 \ln(x) + \frac{x-1}{4}
(x^3-5x^2-2x ) \right] \nonumber \\
&+&   \frac{m_t}{m_b} \krc \frac{1}{2 (x-1)^{3}} \left[
3x\ln(x)+2-\frac{3x}{2}-\frac{x^3}{2}  \right] \,.
\ena
Hence, the $B\to X_s\gamma$ branching ratio predicted by the 
effective theory (\ref{lagrangian}) is:  
\bea
\mbox{Br}( B\to X_s\gamma ) \times 10^{4} =
3.07 + 280 \mbox{Re}(\krc) + 2 \mbox{Re}(\klc)+
5520 {| \krc |}^2  \nonumber \\
+0.3 {| \klc |}^2 + 79 \left( \mbox{Re}(\klc)
\mbox{Re}(\krc) +
\mbox{Im}(\klc) \mbox{Im}(\krc)  \right) \; .
\label{bsgabr}
\ena
It is important to note that the coefficients of
the terms proportional to $\klc$ are at least two orders
of magnitude smaller than their $\krc$ counterparts.
Roughly speaking, only very high values of $\klc$ (of order 1) 
would give a significant contribution.  From the theoretical
standpoint we don't expect such extreme possibility to
occur because new physics effect would likely modify the
anomalous couplings at loop level, hence, at the order of 
$1/4 \pi$ or $1/16 \pi^2$.
For this reason, from now on we will restrict the
possible values of $\klc$ to be at most 0.2, and drop
quadratic terms such as the ones in Eq.~(\ref{bsgabr}).

From the above results, 
we can use the recent result~(\ref{bsgaexp}) from CLEO
to set limits on the real and imaginary  parts of $\krc$ as:
\bea
-.0035 &&\leq \mbox{Re}(\krc) +
20 |\krc |^2 \leq  0.0039 \; ,
\label{eqbsga}
\ena
where, the $2 \sigma$ deviation, that
corresponds closely to $95\%$ CL, was used.
Since the coefficient of the quadratic term, which contains
the contribution from Im($\krc$), is 20 times higher than
that of the single Re($\krc$), we could imagine a case in
which very high values (of order $\pm0.02$, for instance)
of the imaginary part would give a large contribution
which could be counter balanced by another large
and negative contribution from the real part.
Such a situation in which the CP violating coupling would
be one order of magnitude bigger than the CP even real
part is very unusual, though possible.
In Fig.~\ref{krcbounds},
we display the correlated allowed region for Re($\krc$)
and Im($\krc$) defined inside the solid lines.
As to be discussed below, Im($\krc$) can
be better probed with other experimental observables.
As for the information already given by the branching
ratio of $B\to X_s\gamma$,
we conclude that at the $2 \sigma$ level 
\bea
|\mbox{Re}(\krc)| \leq 0.4 \times 10^{-2} \, .
\ena

\subsection{Measuring CP violating couplings}
\label{imagin}
\indent

\begin{figure}
\centerline{\hbox{
\psfig{figure=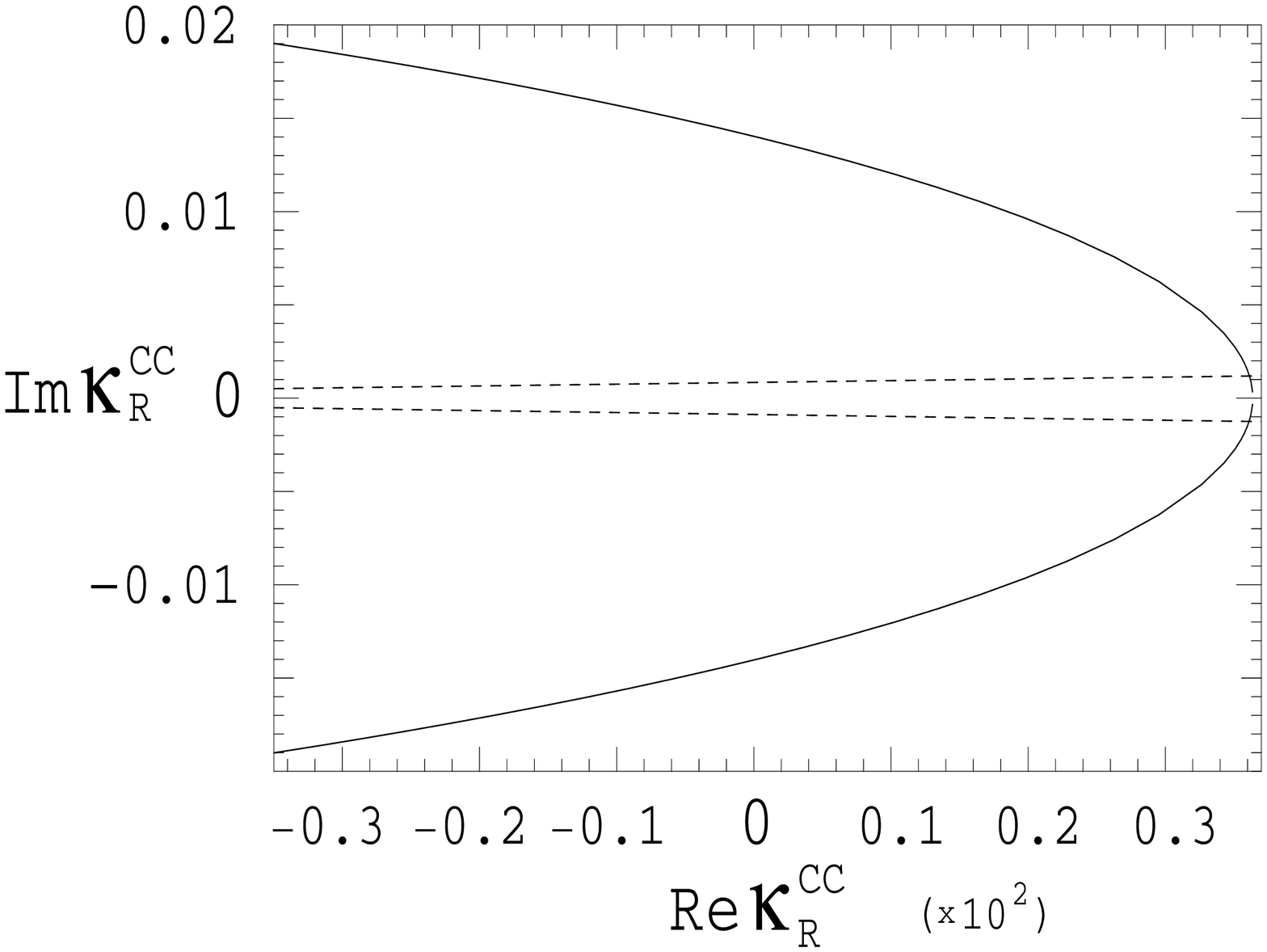,height=7.5in}}}
\caption{Allowed region for $\krc$ from 
the measurement of the branching ratio of $b \to s \gamma$,
between the upper and lower solid line curves.
Region between dashed lines defined for the asymmetry 
$A_{\rm CP}^{b\to s\gamma} \leq 0.25$.}
\label{krcbounds}
\end{figure}

A non-vanishing Im($\krc$) would signal a CP-violation effect.
What do we know about this CP-violating $\tbw$ anomalous
coupling?  So far, there is only one experimental measurement
that gives us some information on Im($\krc$), and that is
the $b\to s\gamma$ branching ratio itself.  As presented in
Eq.~(\ref{bsgabr}) this branching ratio is already sensitive
to a CP violating coupling, and some
constraining region can be already set for Im($\krc$) as is
shown in Fig.~\ref{krcbounds}. 
On the other hand, there can be another observable of
the $b \to s \gamma$ process that can be used to measure CP
violation in the  $\tbw$ coupling.  The following asymmetry has
been proposed to measure CP violation contained in the $C_{2,7,8}$
coefficients \cite{neubert2}:
\begin{eqnarray}
   A_{\rm CP}^{b\to s\gamma}(\delta) &=& \frac{\Gamma(\bar B\to
   X_s\gamma)-\Gamma(B\to X_{\bar s}\gamma)} {\Gamma(\bar B\to
   X_s\gamma)+\Gamma(B\to X_{\bar s}\gamma)}
   \Bigg|_{E_\gamma>(1-\delta) E_\gamma^{\rm max}} \nonumber\\ &=&
  a_{27}(\delta)\,\mbox{Im}\!\left[ \frac{C_2}{C_7} \right] +
   a_{87}(\delta)\,\mbox{Im}\!\left[ \frac{C_8}{C_7} \right] 
 + a_{28}(\delta)\,\frac{\mbox{Im}[C_2 C_8^*]}{|C_7|^2}  \, ,
\label{acp}
\end{eqnarray}
where $C_7$ and $C_8$ are given at scale $m_b$.
As before, there is a dependence on the
energy range of the photon.  
Following Ref.~\cite{neubert2}, we consider the asymmetry for 
$\delta = 0.15$,
then we have $a_{27}=1.31$, $a_{87}=-9.52$,
and $a_{28}=0.07$. 
In terms of the anomalous charged current couplings the
asymmetry reads as follows:
\begin{eqnarray}
   A_{\rm CP}^{b\to s\gamma}(\delta = 0.15) &=&
\left( 31 \, \mbox{Im}(\krc)+0.2 \, \mbox{Im}(\klc) +\right.
\nonumber \\
&& \left. 1.2 \, [\; \mbox{Re}(\klc) \mbox{Im}(\krc)-
\mbox{Re}(\krc)\mbox{Im}(\klc) \;] \right) / |C_7|^2 \, ,
\label{acp2}
\end{eqnarray}
where
\bea
| C_7 |^2 = | C_7 (m_b) |^2 &&=
(19.9 \mbox{Im}(\krc) + 0.141 \mbox{Im}(\klc) )^2 +
\nonumber \\
&& (0.319+19.9 \mbox{Re}(\krc) + 0.141 \mbox{Re}(\klc) )^2
\, .
\ena
Again, we can simplify the above equation by neglecting terms
with $\klc$; here too, the numerical coefficients of $\klc$
are much smaller than those of $\krc$ terms.
We find that 
\begin{eqnarray}
   A_{\rm CP}^{b\to s\gamma}(\delta=0.15) &=&
\frac{\mbox{Im}(\krc)}{0.0031 + 0.41 \mbox{Re}(\krc) +
12.8 |\krc|^2} \, .
\label{asym}
\end{eqnarray}

Notice that indeed this asymmetry is quite sensitive to
$\mbox{Im}(\krc)$ which is consistent with the conclusion of
Ref.~\cite{neubert2} that left-right symmetric models can give
large contribution to the asymmetry $A_{\rm CP}^{b\to s\gamma}$.
As shown in Fig.~\ref{krcbounds},
this asymmetry can set very strong constraints on Im($\krc$).
For instance, if $ A_{\rm CP}^{b\to s\gamma}$ proves to be
smaller than $25\%$, it would mean Im($\krc$) less than $10^{-3}$.

What about Im($\klc$)?  As shown above, the $b\to s \gamma$
process does not make a good probe of the left-handed
CP-odd $\tbw$ coupling. Nevertheless, there are other B-decay processes
with a good potential to measure Im($\klc$) in
future B factories.  For instance, the hadronic channels
$B_d \ra \phi K_s$ and $B_d \ra \Psi K_s$ have been considered
in Ref.~\cite{valencia} for B factories. We shall not discuss
it further in this paper.

\section{Top quark couplings and LEP/SLC data}
\label{lep}
\indent

The validity of the SM at the electroweak loop level has been
established with a very high precision in the recent (and
almost final) results from LEP and SLAC \cite{langacker}.
Except for the Forward-Backward asymmetry ($A^b_{FB}$)
of the b-quark and
the total Left-Right  asymmetry ($A_{LR}$) of $Z \to f \bar f$, there is
a $1\sigma$ or better agreement with the experimental data.
In the light of the remarkable experimental achievement given
by the accuracy of the measurements, and also the
degree of precision in the SM predictions, this agreement would
impose strong limits on the anomalous couplings of the effective
Lagrangian in (\ref{lagrangian}).
In principle, the low energy effective theory can be 
applied to describe an underlying new physics dynamics with or 
without a
Higgs boson, for simplicity, we assume that there exists a SM-like
Higgs boson with mass of 70\,GeV, which brings the
SM predictions to an optimum agreement with the data
\cite{langacker, altarelli}, and concentrate on the effect from the 
anomalous couplings of the top quark.
We first consider all the data that
is consistent with the SM prediction 
within $1 \sigma$, and use them to
constrain the allowed values of the anomalous
$\kappa$ terms in Eq.~(\ref{lagrangian}) at the $2 \sigma$
level. Then, we discuss the possible predictions on $A^b_{FB}$ 
and $A_{LR}$ produced by the constrained $\kappa$'s.

There are two observables of $Z$-pole physics that are particularly
sensitive to top quark couplings as they are proportional to
the top quark mass.  These are the $\rho$ parameter, and
the $b$-$b$-$Z$ vertex; directly associated to $\epsilon_1$ and
$\epsilon_b$ in the analysis by Altarelli,
et. al \cite{altacara}.  The net non-standard contributions
to the $\epsilon$ parameters are

\beq
\delta\epsilon_1=\frac{3{m_t^2} G_F}{2\sqrt{2}{\pi}^2}
 \left( \krn-\kln+\klc-({\krn})^2-({\kln})^2+({\klc})^2+2 \krn
\kln \right)
\ln{\frac{{\Lambda}^2}{m_t^2}}\,\, , \label{cal1}
\enq
\beq
\delta\epsilon_b=\frac{{m_t^2} G_F}{2\sqrt{2}{\pi}^2}
\left ( \kln - \frac{1}{4}\krn \right )
\left ( 1 + 2 \klc  \right )
\ln{\frac{{\Lambda}^2}{m_t^2}}
\,\, , \label{cal2}
\enq
in which only contributions proportional to 
($m_t^2\ln {\Lambda}^{2}$) are kept \cite{malkawi}, 
and the cut-off scale of the
effective theory $\Lambda$ is taken to be  
$4 \pi v \simeq 3$\,TeV \cite{nda}. 
Note that $\klc$ contributes to $\epsilon_b$ 
up to this order only through
the contribution proportional to $\kln$ and $\krn$; since
we want to consider all possible values (within $\pm 0.2$)
of $\klc$ we choose to keep it there.
Given the above results we can then use the experimental values
of the $\epsilon$'s to constrain the theoretical 
predictions \cite{altarelli}:
\bea
1.54 \times 10^{-3} \leq
\epsilon^{{\rm SM}}_1+\delta\epsilon_1
\leq 5.86 \times 10^{-3}
\,\, ,\label{cal11} \\
-8.32 \times 10^{-3} \leq
\epsilon^{{\rm SM}}_b+\delta\epsilon_b
\leq -0.88 \times 10^{-3}
\,\, ,\label{cal21}
\ena
where the minimum and maximum limits represent $2 \sigma$
deviations from the central values of the experimental
measurements.  
From Ref.~\cite{altarelli} we recall the SM values
for $\epsilon_1$ and $\epsilon_b$ are:
$\epsilon^{{\rm SM}}_b = -6.5 \times 10^{-3}$ and
$\epsilon^{{\rm SM}}_1= 5.5 \times 10^{-3}$
for $m_t = 173.8$\,GeV and $m_H=70$\,GeV.

Using the $\kappa$'s contribution as well as the SM values
of $\epsilon_1$ and $\epsilon_b$ given above we
obtain the following inequalities: 
\bea
-0.019 &&\leq  (\krn - \kln) - (\krn - \kln)^2
+ \klc + {\klc}^2  \leq 0.0013
\,\, , \label{bounds1} \\
-0.33 &&\leq (\krn - 4 \kln) \;
(1 + 2 \klc ) \leq 0.1
\,\, . \label{bounds2}  
\ena
Although the above bounds does not take into account the strong
correlation among the possible values of the $\epsilon$'s,
which is described by a 4 dimensional hyperboloid,
it is instructive to find out at this level what is the
implication from these two bounds.
In general, the constraints to the $\rho$ parameter
 imply an almost linear relation:
\bea
\klc \simeq \kln - \krn \, .
\;\;\;\;  \label{krelation}
\ena
The purpose of keeping the quadratic terms in Eq.~(\ref{bounds1})
is to verify that indeed their presence is not significant,
provided we do not consider the highly unlikely possibility
of very big deviations of the top quark couplings (above $20\%$).

\begin{figure}
\centerline{\hbox{
\psfig{figure=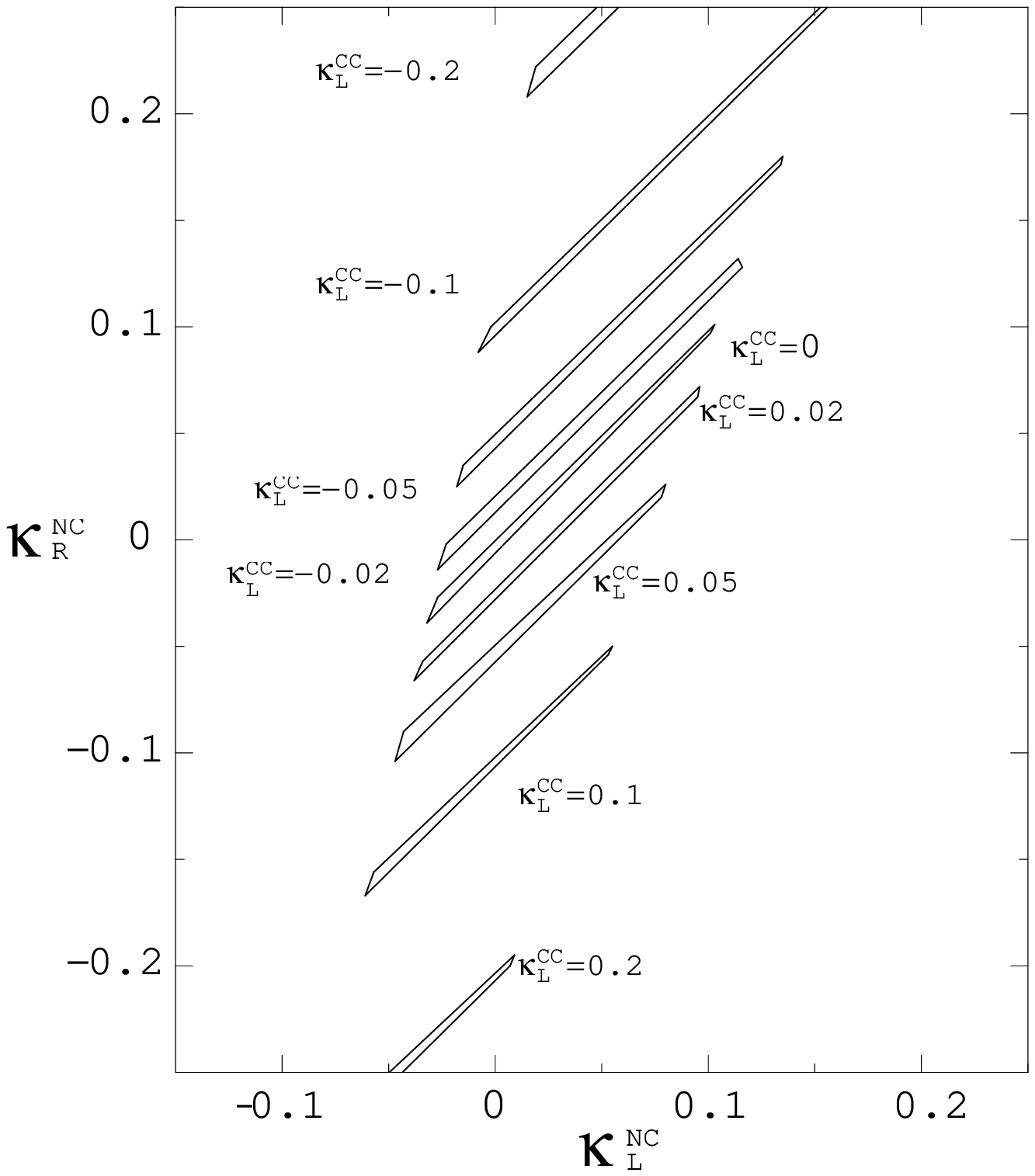,height=5.5in}}}
\caption{Allowed regions for $\krn$ and $\kln$ at different
values of  $\klc$.}
\label{ncbounds}
\end{figure}

To improve the above analysis, we have perform a
$2 \sigma$ fit of the 
$\kappa$'s to the LEP/SLC observables, which includes  
 $\Gamma_Z$, $\sigma_h$, $R_e$, $R_\mu$,
$R_\tau$, $R_b$, $R_c$, $A_{FB} (e)$, $A_{FB} (\mu)$,
$A_{FB} (\tau)$,  $A_{FB} (c)$, $A_{LR} (\mbox{leptons})$,
$A_{\tau} ({\cal P}_\tau)$ and  $A_{e} ({\cal P}_\tau)$
\cite{langacker},
as well as the $m_W^2/m_Z^2$ ratio \cite{altarelli}.
We find that the linear relation of
Eq.~(\ref{krelation}) is now very precisely established,
as the remarkably narrow allowed regions of $\krn$ and $\kln$
shown in Fig.~\ref{ncbounds}
sharply describe segments of lines with slope equal to 1.
Therefore, we conclude that the LEP/SLC data have strongly
constrained the anomalous 
couplings of Eq.~(\ref{lagrangian}).
Even though one of them, $\klc$ for instance,
can take on any value (within 0.2), the two other couplings $\krn$
and $\kln$ are constrained to satisfy Eq.~(\ref{krelation})
and cannot be far off from the value of $\klc$ itself. 

\begin{figure}
\centerline{\hbox{
\psfig{figure=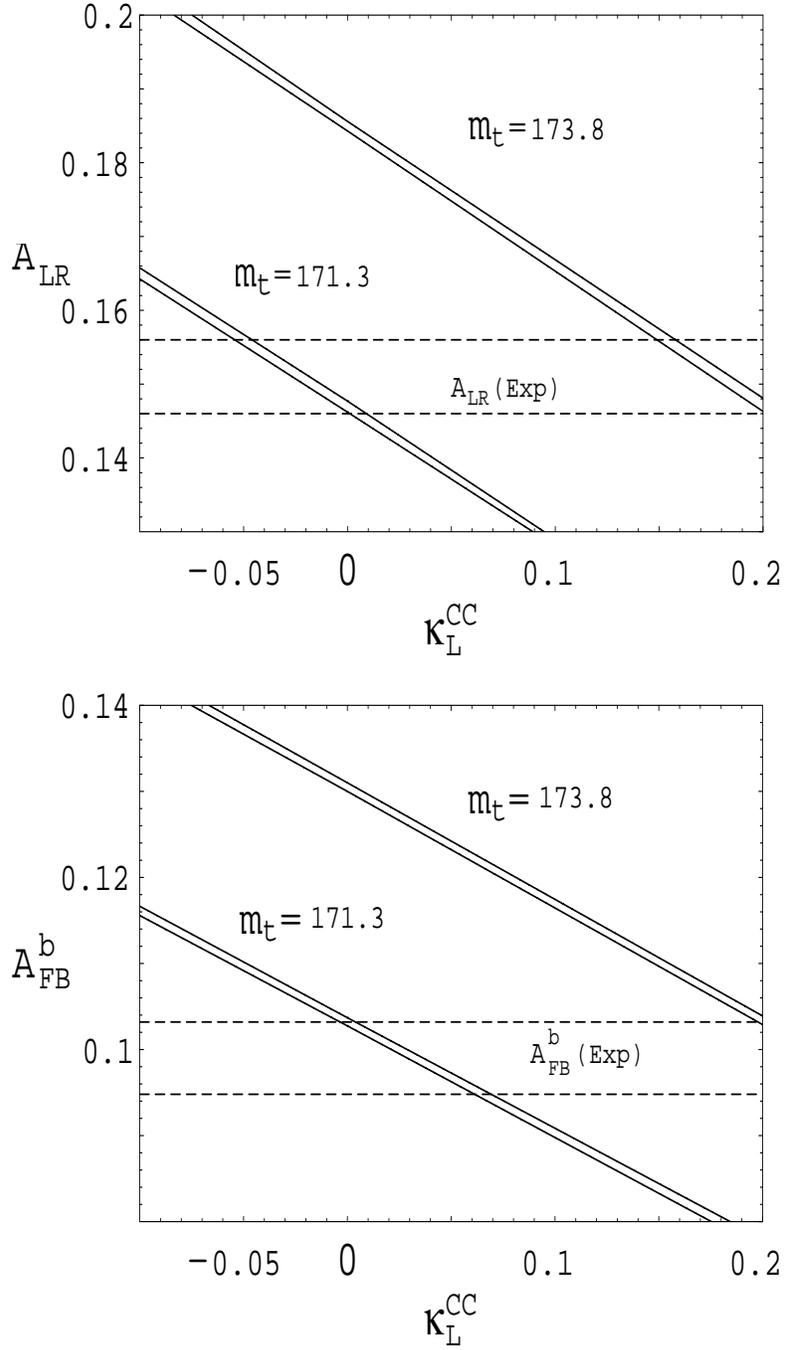,height=7.5in}}}
\caption{Asymmetries of $Z$ decays as a function of $\klc$.  The
narrow regions between solid lines come from the small possible
variation of the correlated $\krn$ and $\kln$ couplings.}
\label{assym}
\end{figure}

Let us now consider the three LEP/SLC observables that show
an almost $2 \sigma$ deviation from the SM, namely  $A^b_{FB}$,
$A_{LR}$(hadrons) and $A_b$ \cite{langacker}.
Could the $\tbw$ and $\ttz$ anomalous couplings account for
these discrepancies?
In Fig.~\ref{assym} we show the predicted possible values of
$A_{LR}$ and $A^b_{FB}$ for the two values of $m_t$.
%cpy
(Recall that 
the CDF/D0 direct measurement gives \cite{cdfd0}
$
m_t^{CDF/D0} = 173.8 \pm 5.0  \; \mbox{GeV} \; ,
$ 
which combined with LEP/SLC data gives \cite{langacker,altarelli} 
$
m_t^{fitted} = 171.3 \pm 4.9 \; \mbox{GeV} \; .
$)
The $2 \sigma$ experimental range of these asymmetries are
shown by the dashed lines.  For a given $m_t$, 
the solid lines define a very
narrow region of predicted values coming from the allowed
values of the $\kappa$ couplings.
For $m_t = 171.3$ GeV, the $A_{LR}$ data prefers 
a negative $\klc \sim -0.04$, whereas $A_{FB}^b$ favors
a similar but positive value of $\klc$.
On the other hand,
for $m_t = 173.8$ GeV, $A_{LR}$ would require
$\klc \sim 0.15$, but again $A^b_{FB}$ would need a different
value of $\klc$ (bigger than 0.2 in this case).
If it turns out that $m_t$ is about 172 GeV, then a value
of $\klc \sim 0.1$ could explain both $A_{LR}$ and
$A_{FB}^{b}$.  Such a value of $\klc$ would not modify the
$M_W$ dependence on $m_t$ given by the SM.\footnote{We have
also checked that, for $\mid \klc \mid < 0.2$, the
correlation between $M_W$ and $m_t$ is almost identical
to the SM prediction.}

\section{Conclusions}
\label{conclu}
\indent

Inspired by the fact that no satisfactory proven mechanism for
the breaking of the electroweak symmetry exists, and the fact
that the top quark stands out as much heavier than all the
other known elementary particles, we proposed a model in which
the $\ttz$ and $\tbw$ couplings depart from their SM values. 
The SM has been highly tested by the reasonably good measurements
of rare decay processes, such as $b\to s \gamma$, by the
CLEO collaboration, as well as by
the precision data of LEP and SLC.  This is reflected in the
very stringent constraints on the top quark couplings. 
 For instance,
the measurement of $b\to s \gamma$ alone sets a constraint of
less than $0.5 \%$ for the possible strength of a right handed
$\tbw$ coupling (in terms of the SM $g/{\sqrt{2}}$ value) and
even a $2 \%$ upper limit for the size of an imaginary CP odd
part.   On the other hand, the LEP/SLC data (even though they do
not restrict all the anomalous $\kappa$ terms since possible 
cancellations are allowed), impose strong correlations on 
$\kappa$'s so that if
only one coupling, $\klc$ for instance, is not zero, the others
are forced to be of about the same order of magnitude.
Also, given this strong correlation of the anomalous couplings, one
can find out their precise prediction of the forward-backward 
$b$ quark asymmetries of $Z$ decay
and left-right asymmetry $A_{LR}$.
  At present, there is a $1.8 \sigma$ deviation
of the experimental measurement from the SM prediction.  It turns out
that these anomalous couplings have little chance of predicting
such discrepancy.  To do so, two requisites have to be met:
({\it i}) $\klc$ has to be of order 0.1, and 
({\it ii}) the
mass of the top quark should prove to be nearly 172 GeV.
Fortunately, at the Run-2 of the Tevatron, it would be possible
to measure $\klc$ to $\sim 5\%$ accuracy via measuring the
single-top production rate \cite{tait}.

\medskip
\section*{ \bf  Acknowledgments }
\indent \indent

We thank H.-J. He, E. Malkawi, T. Tait and V. Gupta for
helpful discussion. 
F.L. would like to thank Conacyt for support.
C.P.Y. was supported in part by the NSF grant
PHY-9802564.
We also thank the Aspen Center for Physics for their hospitality.
Part of this work was funded by the joint NSF-Conacyt project:
{\it Tests of SUSY and Electroweak Chiral Lagrangians}.

\end{document}